\documentclass[final]{aa}

\pdfoutput=1

\usepackage{graphicx}
\usepackage{txfonts}
\usepackage{natbib}
\bibpunct{(}{)}{;}{a}{}{,} 

\def\solar {\ifmmode_{\mathord\odot} \else $_{\mathord\odot}$ \fi}
\def\jup {\ifmmode_{\mathrm{Jup}} \else $_{\mathrm{Jup}}$ \fi}
\def\earth {\ifmmode_{\mathord\oplus} \else $_{\mathord\oplus}$ \fi}
\def\Msol {\ifmmode {\,\mathrm{M}\solar} \else \,M\solar \fi}     
\def\Rsol {\ifmmode {\,\mathrm{R}\solar} \else R\solar \fi}     
\def\Lsol {\ifmmode {\,\mathrm{L}\solar} \else L\solar \fi}     
\def\Mjup {\ifmmode {\,\mathrm{M}\jup} \else M\jup \fi}
\def\Mearth {\ifmmode {\,\mathrm{M}\earth} \else M\earth \fi}

\def\mps {\ifmmode {\,\mathrm{m\,s^{-1}}} \else $\mathrm{m\,s^{-1}}$ \fi}     

\begin{document}
  \title{The HARPS search for southern extra-solar planets\thanks{Based on observations made with the HARPS instrument on the ESO 3.6 m telescope under the GTO program ID 072.C-0488 at Cerro La Silla (Chile).}}

  \subtitle{X. A $m \sin i = 11 \Mearth$ planet around the nearby spotted M dwarf \object{GJ~674}}

\author{X.~Bonfils   \inst{1} 
   \and M.~Mayor     \inst{2}
   \and X.~Delfosse \inst{3}
   \and T.~Forveille  \inst{3}
   \and M.~Gillon   \inst{2}
   \and C.~Perrier     \inst{3}
   \and S.~Udry         \inst{2}
   \and F.~Bouchy    \inst{4}
   \and C.~Lovis       \inst{2}
   \and F.~Pepe        \inst{2}
   \and D.~Queloz    \inst{2}
   \and N.~C.~Santos \inst{1,2,5}
   \and J.-L.~Bertaux \inst{6} 
}

  \offprints{X. Bonfils}
  \institute{
Centro de Astronomia e Astrof{\'\i}sica da Universidade de Lisboa,
    Observat\'orio Astron\'omico de Lisboa, Tapada da Ajuda, 1349-018
    Lisboa, Portugal
  \email{xavier.bonfils@oal.ul.pt}
  \and
  Observatoire de Gen\`eve, 51 ch. des Maillettes, CH-1290 Sauverny, Switzerland
  \and
  Laboratoire d'Astrophysique, Observatoire de Grenoble, BP 53, F-38041 Grenoble, Cedex 9, France
  \and
  Institut d'Astrophysique de Paris, CNRS, Universit\'e Pierre et Marie Curie, 98bis Bd Arago, 75014 Paris, France
    \and
    Centro de Geofisica de \'Evora, Rua Rom\~ao Ramalho 59,
    7002-554 \'Evora, Portugal
   \and
  Service d'A\'eronomie du CNRS, BP 3, 91371 Verri\`eres-le-Buisson, France
            }

  \date{Received January 09, 2007; accepted xxxx xx, 2007}


 \abstract
  {How planet properties depend on stellar mass is a key diagnostic of planetary formation
   mechanisms.
  }
  {This motivates planet searches around stars which are significantly more massive or less
   massive than the Sun, and in particular our radial velocity search for planets around
   very-low mass stars.
  }
  {
  As part of that program, we obtained measurements of
 \object{GJ~674}, an M2.5 dwarf at d=4.5~pc, which have a
  dispersion much in excess of their internal errors.
  An intensive observing campaign demonstrates that the
  excess dispersion is due to two superimposed coherent
  signals, with periods of 4.69 and 35 days.
  }
  {These data are well described by a 2-planet
  Keplerian model where each planet has a
  $\sim$11 \Mearth minimum mass. A careful analysis of
  the (low level) magnetic activity of
  \object{GJ~674} however demonstrates that the 35-day
  period coincides with the stellar  rotation period.
  This signal therefore originates in a spot inhomogeneity
  modulated by stellar rotation. The 4.69-day signal on the
  other hand is caused by a bona-fide planet,
  \object{GJ~674b}. %
  }
  {Its detection adds to the growing number of
   Neptune-mass planets around M-dwarfs,
   and reinforces the emerging conclusion that this
   mass domain is much more populated
   than the jovian mass range. We discuss the metallicity 
   distributions of M dwarf \textit{with} and \textit{without} 
   planets and find a low 11\% probability that they are drawn 
   from the same parent distribution. Moreover, we find tentative
   evidence that the host star metallicity correlates with the 
   total mass of their planetary system.}


\titlerunning{An 11$\Mearth$ planet around the nearby M dwarf \object{GJ~674}}
\authorrunning{X. Bonfils et al.}

  \maketitle


\section{Introduction}
M dwarfs, the most common stars in our Galaxy,
were added to the target lists of planet-search
programs soon after the first exoplanet discoveries.
Compared to Sun-like stars, they suffer from some
drawbacks: they are faint and photon noise therefore
often limits measurements of their radial velocity,
and many are at least moderately active and thus
prone to so-called ``radial-velocity jitter'' 
\citep{Saar1997}. On the
other hand, the smaller masses of M dwarfs result in a
higher wobble amplitude for a given planetary mass,
and their p-mode oscillations have both smaller
amplitudes and shorter periods than those of solar
type stars. These oscillations therefore average 
out much faster. As
a result, the detection of an Earth-like planet in
the -- closer -- habitable zone of an M dwarf is
actually within reach of  today's best spectrographes.
Perhaps most importantly, however, M dwarfs represent
unique targets to probe the dependance on stellar mass
of planetary formation, thanks to the wide mass range
(0.1 to 0.6$\Msol$) spanned by that spectral class alone.

The first planet found to orbit an M dwarf,
\object{GJ 876}b \citep{Delfosse1998b, Marcy1998},
was only the 9$^{th}$ exoplanet discovered around a main
sequence star. Besides showing that Jupiter-mass
planets can form at all around very-low-mass stars,
its discovery suggested that they might be common,
since it was found amongst the few dozen M dwarfs that were
observed at that time. Against these early expectations,
no other M dwarf was reported to host a planet until 2004,
though a second planet 
\citep[\object{GJ 876c}, $m_p~\sin~i~=~0.56~\Mjup$ --][]{Marcy2001} was
soon found around \object{GJ 876} itself.

In 2004, the continuous improvement of the
radial-velocity techniques resulted in the
quasi-simultaneous discovery of three Neptune-mass planets,
around \object{$\mu$ Ara}
\citep[$m_p \sin i = 14 \Mearth$ --][]{Santos2004},
\object{$\rho$ Cnc}
\citep[$m_p \sin i = 14 \Mearth$ --][]{McArthur2004} and
\object{GJ 436}
\citep[$m_p \sin i = 23 \Mearth$ --][]{Butler2004, Maness2006}.
Of those three, \object{GJ 436b}, orbits an M dwarf,
and put that spectral class back on the discovery forefront.
It was soon followed by another two, a single planet around
\object{GJ 581} \citep[$m_p \sin i = 17 \Mearth$ --][]{Bonfils2005b} and a
very light ($m_p = 7.5 \Mearth$) third planet in the \object{GJ 876}
system \citep{Rivera2005}. As a result, planets around M dwarfs
today represent a substantial fraction (30\%)
of all known planets with $m\sin i \la 30 \Mearth$.

Even with \object{GJ 849}b
\citep[$m_p \sin i= 0.82 \Mjup$ --][]{Butler2006} now
completing the inventory of M-dwarf planets found with
radial-velocity techniques, the upper-range of planet
masses remains scarcely populated. This contrasts both with
the (still very incompletely known) Neptune-mass planets
orbiting M dwarfs and with the jovian planets around
Sun-like stars. At larger separations, microlensing
surveys similarly probe the frequency of planets as a
function of their mass. That technique has detected four
putative planets that likely orbit M dwarfs:
\object{OGLE235-MOA53b}
\citep[$m_p\sim1.5-2.5\Mjup$ --][]{Bond2004},
\object{OGLE-05-071Lb}
\citep[$m_p=0.9\Mjup$ --][]{Udalski2005},
\object{OGLE-05-390Lb}
\citep[$m_p=0.017\Mjup$ --][]{Beaulieu2006}
and \object{OGLE-05-169Lb}
\citep[$m_p=0.04\Mjup$ --][]{Gould2006}. Two of
these four planets have likely masses
below 0.1~\Mjup. Given the detection bias of that
technique towards massive companions, this again
suggests that Neptune-mass planets are
much more common than Jupiter-mass ones
around very-low-mass stars.

Here we report the discovery of a 11 \Mearth
planet orbiting \object{GJ~674} every 4.69~days. \object{GJ~674b}
has the 5$^{th}$ lowest mass of the known planets, and coincidentally is also
the 5$^{th}$ planetary system centered on a M dwarf. Its detection
adds to the small inventory of both very-low mass planets and planets
around very-low mass stars. After reviewing the properties of
the \object{GJ 674} star (\S\ref{sect:prop}), we briefly present
our radial velocity measurements (\S\ref{sect:data}) and
their Keplerian analysis (\S\ref{sect:orbit}). A careful
analysis of the magnetic activity of \object{GJ 674}
(\S\ref{sect:activity}) assigns one of the two periodicities
to rotational modulation of a stellar spot signal, and
the other one to a {\it bona fide} planet. We
conclude with a brief discussion of the properties of the
detected planet.

\section{\label{sect:prop}The properties of \object{GJ~674}}
\object{GJ~674} (\object{HIP 85523}, \object{LHS 449})
is a M2.5 dwarf \citep{Hawley1997} in the Altar constellation.
At $4.5~\mathrm{pc}$
\citep[$\pi = 220.43 \pm 1.63~\mathrm{mas}$ --][]{ESA1997},
it is the 37th closest stellar system, the 54th closest star
(taking stellar multiplicity into
account)\footnote{on Mar. 1$^{st}$ 2007
(http://www.chara.gsu.edu/RECONS/TOP100.htm)}, and
only the 2nd closest known planetary system (after
$\epsilon$ Eridani, and slightly closer than
\object{GJ 876}).\  

Its photometry \citep[$V =  9.382 \pm 0.012$;
$K = 4.855 \pm 0.018$ --][]{Turon1993, Cutri2003} and
parallax imply absolute magnitudes of
$M_V = 11.09 \pm 0.04$ and $M_K = 6.57 \pm 0.04$.
\object{GJ 674}'s $J-K$ color 
\citep[$=0.86$ --][]{Cutri2003} and 
the \citet{Leggett2001} colour-bolometric relation result 
in a K-band bolometric correction of $BC_K=2.67$, and 
in a 0.016 \Lsol luminosity.

The K-band mass-luminosity relation of
\cite{Delfosse2000} gives a $0.35 \Msol$
mass and the \citet{Bonfils2005a} photometric
calibration of the metallicity
results in $[\mathrm{Fe/H}]=-0.28\pm0.2$.

\begin{figure}
\centering
\includegraphics[width=0.9\linewidth]{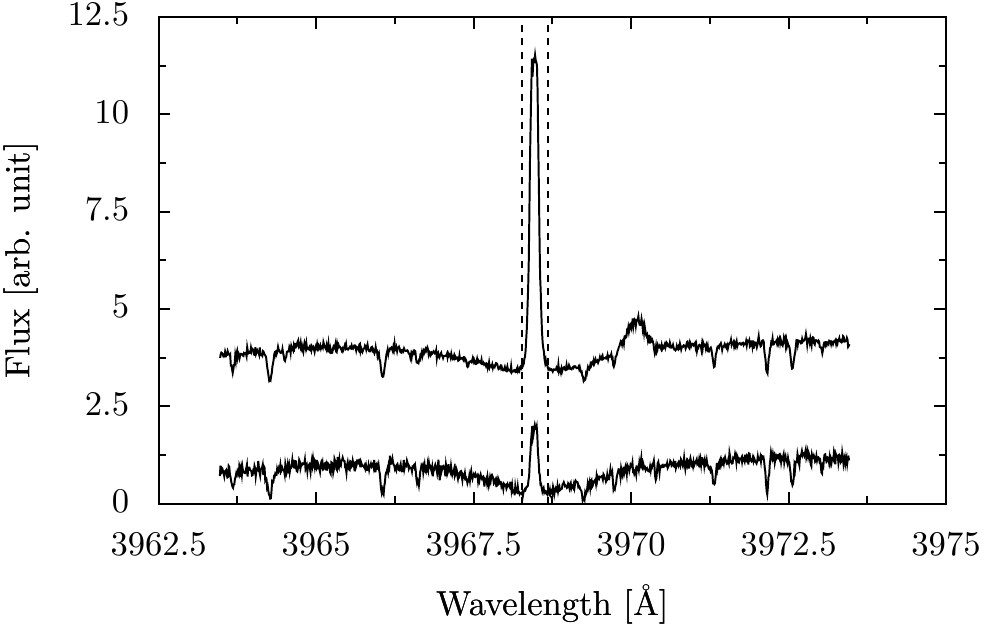}
       \caption{\label{fig:CaIIH}Emission reversal in the 
       \ion{Ca}{ii} H line of \object{GJ~674} (top)
       and \object{GJ 581} (bottom). Within our sample 
       \object{GJ 581} has one of the weakest \ion{Ca}{ii} 
       emission and illustrates a very quiet M dwarf.
       \object{GJ~674} has much stronger emission and is 
       moderately active.
       }
       \label{fig:caii}
\end{figure}

The moderate X-ray luminosity 
\citep[$L_x/L_{bol}\simeq5.10^{-5}$ --][]{Hunsch1999} and
\ion{Ca}{ii} H \& K emission depict a modestly
active M dwarf (Fig.~\ref{fig:CaIIH}). Its UVW galactic velocities place
\object{GJ~674} between the young and old disk populations
\citep{Leggett1992}, suggesting an age of $\sim10^{8-9} \mathrm{yr}$.

Last but not least, since we are concerned with
radial velocities, the high proper motion of
\object{GJ~674}
\citep[$1.05~\mathrm{arcsec\,yr^{-1}}$ --][]{ESA1997}
changes the orientation of its velocity vector
along the line-of-sight \citep[e.g. ][]{Kurster2003}
to result in an apparent secular acceleration of
$0.115 \mps\,\mathrm{yr^{-1}}$. At our current precision
this acceleration will not be detectable before another decade.

\begin{table}
\centering
\caption{
\label{table:stellar}
Observed and inferred stellar parameters for GJ~674}
\begin{tabular}{l@{}lc}
\hline
 \multicolumn{2}{l}{\bf Parameter}
& \multicolumn{1}{c}{\bf GJ~674} \\
\hline
Spectral Type   &                & M2.5\\
V                       &               & $9.382 \pm 0.012$ \\
$\pi$           &[mas]          & $220.43 \pm 1.63$ \\
Distance                &[pc]           & $4.54 \pm 0.03  $\\
$M_V$           &               & $11.09 \pm 0.04$ \\
K                       &               & $4.855 \pm 0.018$\\
$M_K$           &               & $6.57 \pm 0.04 $\\
$L_\star$       & [$\mathrm{L_\odot}$]          &  $0.016$\\
$L_x/L_{bol}$   &               &  $5.10^{-5}$\\
$v\sin i$               & [km\,s$^{-1}$] & $ \lesssim 1 $ \\
$dv_r/dt$              & [m\,s$^{-1}$yr$^{-1}$] & 0.115\\
$[Fe/H]$                &               & $ -0.28 $\\
$M_\star$       & [$\Msol$]             & $ 0.35 $\\
age & [Gyr] & 0.1-1  \\
$T_{\rm eff}$   & [K]           &   3500-3700 \\
\hline
\end{tabular}
\end{table}

\section{\label{sect:data}Radial-velocity data}
We observed \object{GJ~674} with the HARPS echelle
spectrograph \citep{Mayor2003} mounted on the ESO
3.6-m telescope at La Silla Observatory (Chile). After
demonstrating impressive planet finding capabilities
right after its commissioning \citep{Pepe2004},
this spectrograph now defines the state of the art
in radial-velocity  measurements, delivering a
significantly better precision than its ambitious
$1 \mps$ specification. As one recent published example,
\citet{Lovis2006} obtained a $0.64 \mps$ dispersion
for the residuals of their orbital solution of the
3~Neptune-mass planets of \object{HD 69830}.

We observed \object{GJ~674} without interlaced
Thorium-Argon light to obtain cleaner spectra
for spectroscopic analysis, at some small cost
in the ultimate Doppler precision. Since June 2004
we have gathered 32 exposures of 900~s each with a median
S/N ratio of $\sim 90$. Their Doppler information
content, evaluated according to the prescriptions of
\citet{Bouchy2001}, is mostly below 1 $\mps$. 
Our internal errors additionally include, in quadrature sum,
an ``instrumental'' uncertainty of $0.5 \mps$ for the nightly drift of the
spectrograph (since we do not use the ThAr lamp to monitor
it) and the measurements uncertainty of the daily wavelength 
zero point calibration. We did benefit of the recent improvements 
of the HARPS wavelength calibration, which is now stable to $0.1 \mps$ 
\citep{Lovis2006b}.

\begin{figure}
\centering
\includegraphics[width=0.9\linewidth]{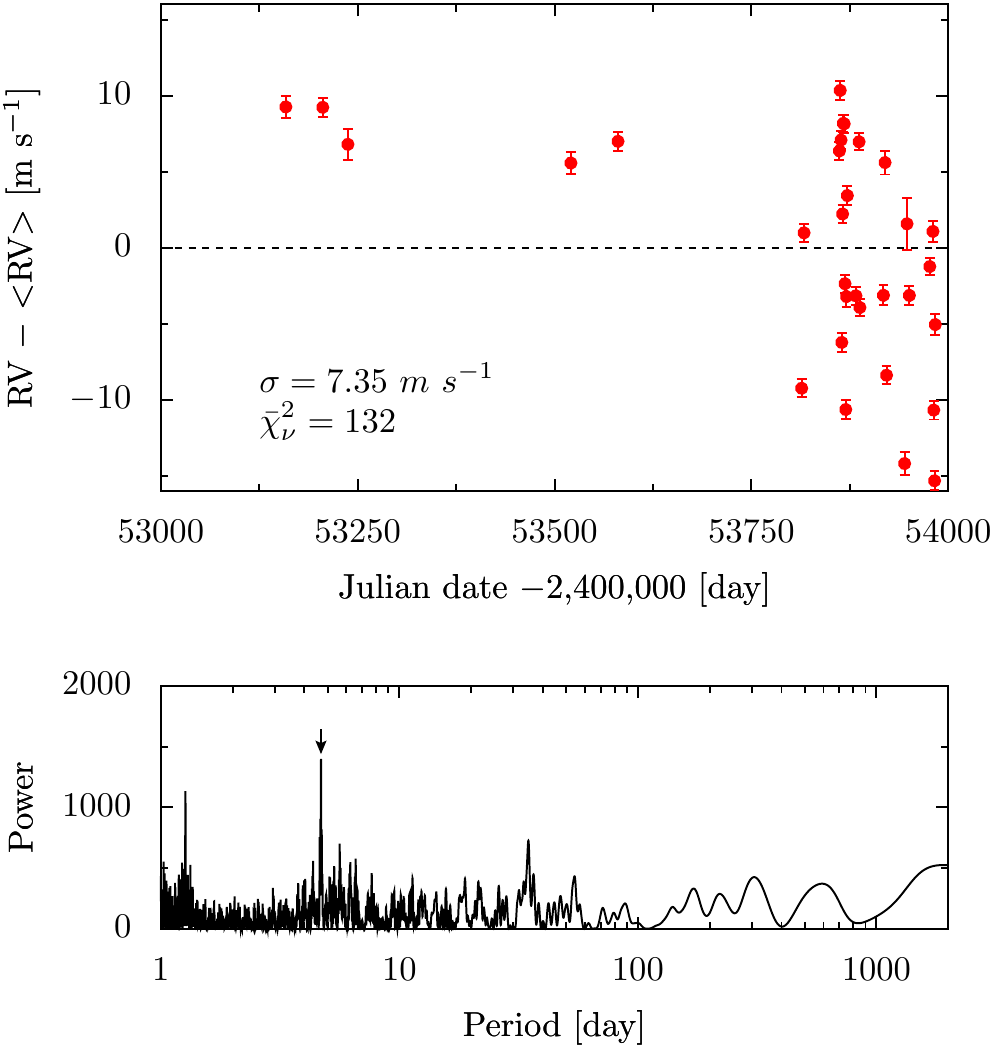}
       \caption{{\it Upper panel:} Radial-velocity 
      measurements of \object{GJ~674} as a function of time.
      The high dispersion ($\sigma = 7.35 \mps$) and chi-square 
      value ($\bar{\chi}^2 = 132$) betray a (coherent or incoherent) 
      signal in the data. {\it Bottom panel:} the Lomb-Scargle periodogram
      of the velocities has prominent power excess around 
      $P = 4.69$~days (downward arrow), which indicates that much 
      of the excess dispersion reflects a coherent signal with a 
      period close to that value. The second highest peak, at 
      1.27~day, is a one-day alias of the 4.69~days period 
      (1.27 $\simeq 1+1/4.69$).}
       \label{fig2}
\end{figure}

A constant radial velocity gives a very large reduced 
chi-square ($\bar{\chi}^2 = 132$) for the time series,
which reflects a dispersion ($\sim 7.4 \mps$) well
above our internal errors (Fig.~\ref{fig2}). This prompted 
a search for an orbital (\S\ref{sect:orbit}) and/or magnetic
activity (\S\ref{sect:activity}) signal.

\section{\label{sect:orbit}Orbital analysis}

\begin{figure}
\centering
\includegraphics[width=0.9\linewidth]{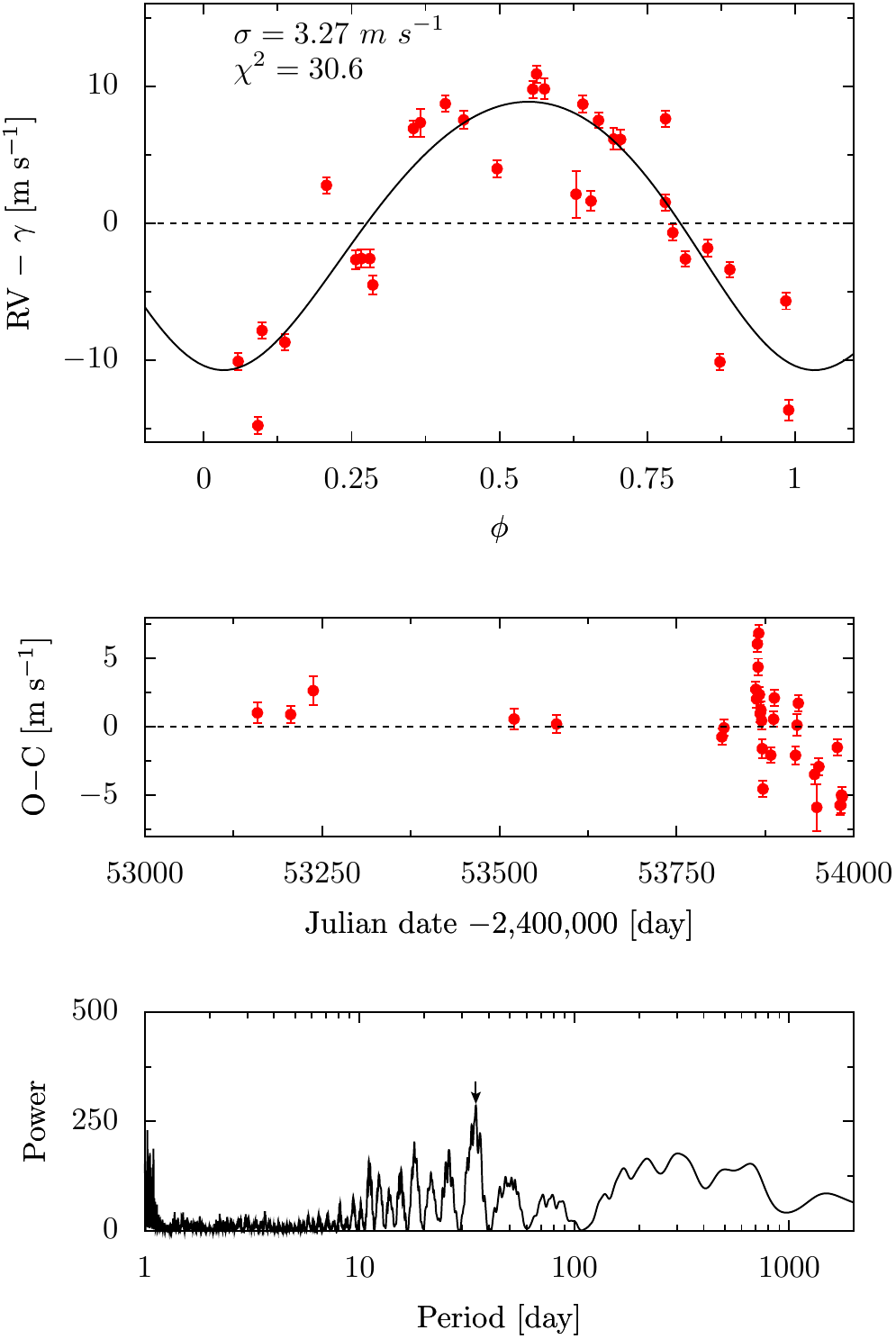}
       \caption{{\it Upper panel:} Radial velocities of 
         \object{GJ~674} (red filled circles) phase-folded to 
         the 4.6940~days period of the best 1-planet fit (curve). 
         The dispersion around the fit ($\sigma = 3.27 \mps$) 
         and its reduced chi-square ($\bar{\chi}^2 = 30.6$ per
         degree of freedom) indicate that a single planet does 
         not describe the data very well.
       {\it Middle panel:} Radial-velocity residuals of the  
         1-planet fit
       {\it Bottom panel:} The Lomb-Scargle periodogram of 
         the residuals shows a broad  peak centered around 
         35 days.}
       \label{fig3}
\end{figure}

A Lomb-Scargle periodogram \citep{Press1992} of
the velocity measurements shows a narrow peak around
4.69-day  (Fig.~\ref{fig2}). Adjustment of a single
Keplerian orbit demonstrates that it is best described
by a $m_2 \sin i = $12.7~\Mearth planet ($0.040~\Mjup$)
revolving around \object{GJ~674} every
$P_2 = 4.6940 \pm 0.0005$ days in a slightly eccentric
orbit ($e_2 = 0.10 \pm 0.02$). The residuals around
this low-amplitude orbit ($K_1 = 9.8 \pm 0.2 \mps$)
have a dispersion of 3.27 \mps (Fig.~\ref{fig3}), still
well above our measurement errors, and the reduced
chi-square per degree of freedom is
$\bar{\chi}^2 = 30.6$. A periodogram of the residuals
indicates that much of this excess dispersion stems from
a broad power peak centered around 35~days, prompting us to
perform a 2-planet fit.

\begin{figure}
\centering
\includegraphics[width=0.9\linewidth]{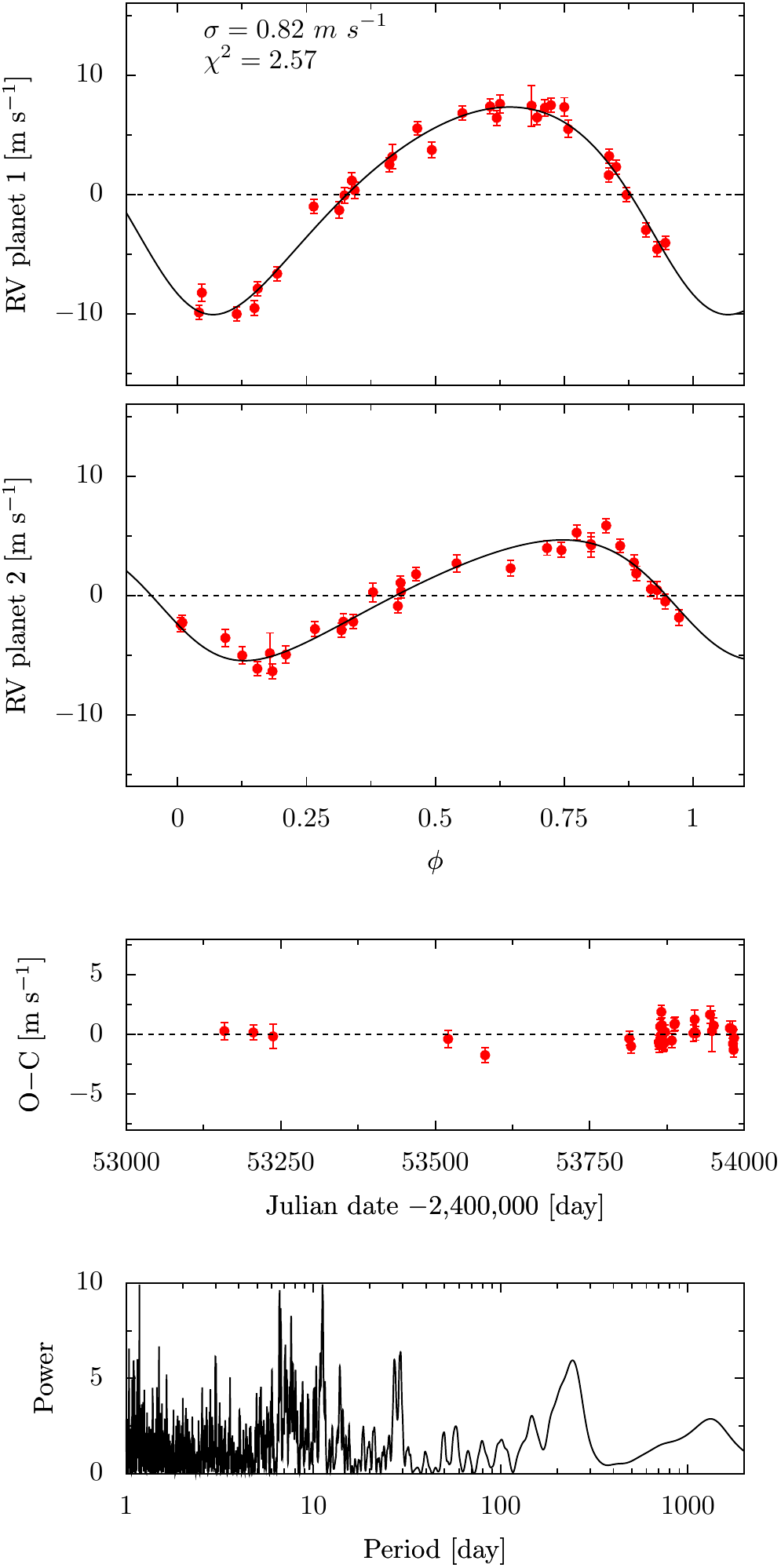}
       \caption{
 {\it Top two panels:} Radial velocity measurements phased to each
  of the two periods, after subtraction of the other component of 
  our best 2-planet Keplerian model. 
{\it Third panel:} Residuals of the best 2-planet fit as a function 
  of time (O$-$C, Observed minus Computed).
{\it Bottom panel:} Lomb-Scargle periodogram of these residuals.
}
       \label{fig4}
\end{figure}

We searched for 2-planet Keplerian solutions with
{\it Stakanof} (Tamuz, in prep.), a program
which uses genetic algorithms to efficiently
explore the large parameter space of multi-planet
models. {\it Stakanof} quickly converged to a
2-planet solution that describes our measurements
much better than the single planet fit ($\sigma = 0.82 \mps$,
$\bar{\chi}^2 = 2.57$ per degree of freedom --  Fig.~\ref{fig4}).
The orbital parameters of the 4.69-day planet change
little from the 1-planet fit, except for the
eccentricity which increases to $e_2 = 0.20 \pm 0.02$. Its
mass is revised down to $M_2 \sin i = 11.09 \Mearth$, and
the period hardly changes, $P_2 = 4.6938 \pm 0.0007$ day.
The second planet would have a $P_3~=~34.8467 \pm 0.0324$ day
period, an $e_3~=~0.20 \pm 0.05$ eccentricity and a minimum 
mass of $m_3 \sin i = 12.58 \Mearth$. Such periods would 
correspond to semi-major axes of 0.04 and 0.15 AU. Those
are sufficiently disjoint that mutual interactions can be 
neglected over observable time scales, and that the system 
would be stable over longer time scales.

The low dispersion around the solution and the lack of any
significant peak in the Lomb-Scargle periodogram of its 
residuals shows that our current radial-velocity 
measurements contain no evidence for an additional
component. 

\section{\label{sect:activity}Activity analysis}
Apparent Doppler shifts unfortunately do not always originate
in the gravitational pull of a companion: in a rotating
star, stellar surface inhomogeneities such as plages and
spots can break the exact balance between light emitted in
the red-shifted and blue-shifted halves of the star. Observationally,
these inhomogeneities translate into flux variations as well
as into changes of both the shape and the centroid of spectral lines
\citep{Saar1997, Queloz2001}. Spots typically also impact spectral
indices, whether designed to probe the chromosphere (to
which photospheric spots have strong magnetic connections), or
the photosphere (because spots have cooler spectra). Of the
two candidate periods, the 4.69-day one is unlikely to reflect
stellar rotation. We measure from our \object{GJ~674} spectra
a rotational velocity of $v\sin i \lesssim 1~\mathrm{km\,s^{-1}}$,
which would need a rather unprobable stellar inclination
($i\lesssim 15\degr$) to match such a short period.
The moderate activity level of \object{GJ~674} on the other
hand leaves the nature of the second signal a priori
uncertain, and the very small rotation velocity removes
much of the power of the usual bisector test 
(Appendix~\ref{subsect:bis}). We therefore investigated 
its magnetic activity through photometric observations 
(\S\ref{subsect:phot}) and detailed examination of the 
chromospheric features in the clean HARPS spectra 
(\S\ref{subsect:spec}).

\subsection{\label{subsect:phot}Photometric variability}

\begin{figure}
\centering
\includegraphics[width=0.9\linewidth]{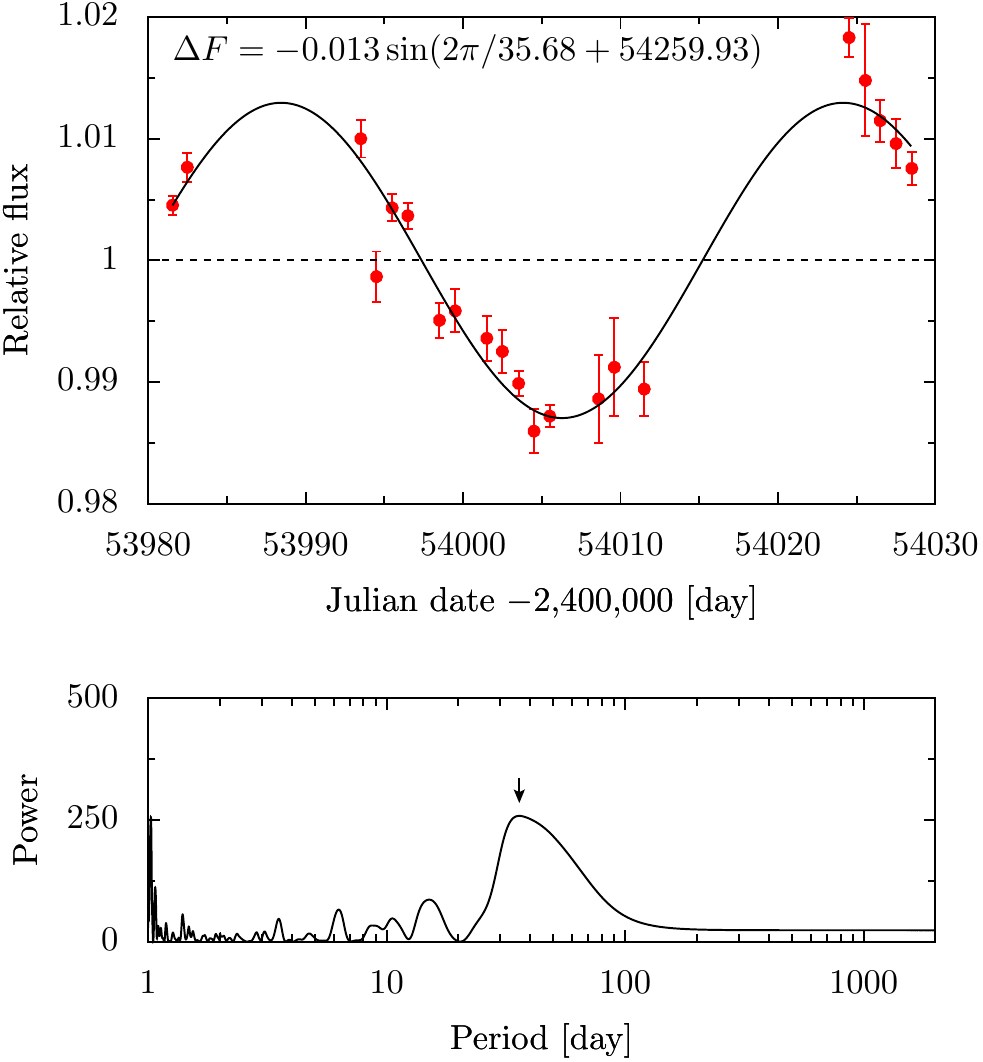}
       \caption{{\it Upper panel:} Differential
       photometry of \object{GJ~674} as a
       function of time. The star clearly varies with
       a 1.3\% amplitude.
       {\it Bottom panel:} The periodogram of the
      \object{GJ~674} photometry exhibits significant
       power excess peaked at 35 days (small black arrow).
       }
       \label{fig:phot1}
\end{figure}

We obtained photometric measurements with
the CCD camera of the Euler Telescope (La Silla)
during 21 nights between September 2$^{nd}$ and October
19$^{th}$ 2006. \object{GJ~674} was observed through a VG filter
which, amongst the available filters, optimizes the
flux ratio between \object{GJ~674} and its two brightest
reference stars. This relatively blue filters also
happens to have good sensitivity to spots on cool stars
such as \object{GJ~674}. To minimize atmospheric scintillation
noise we took advantage of the low stellar density to
defocus the images to FWHM $\sim 11\arcsec$, so that we could use
longer exposure times. The increased read-out and sky
background noises from the larger synthetic aperture which 
we then had to use remain negligible compared to both 
stellar photon noise and scintillation.

We gathered 14 to 75 images per night with a
median exposure time of 20 seconds. We used the
Sept. 24$^{th}$ data, which have the longest nightly
time base,
to tune the parameters of the \textsc{Iraf Daophot}
package and optimize the set of reference stars
(\object{HD 157931}, \object{CD 4611534} and
7~anonymous fainter stars) to minimize the
dispersion in the \object{GJ~674} photometry for
that night. These parameters were then fixed for
the analysis of the full data set. The nightly
light curves for \object{GJ~674} were normalized by
that of the sum of the references, clipped at
3-$\sigma$ to remove a small number of outliers,
and averaged to one measurement per night to examine
the long term photometric variability of \object{GJ~674}.
\object{GJ~674} clearly varies with a $\sim$1.3\% amplitude,
and a (quasi-)period close to 35 days
(Fig.~\ref{fig:phot1}). To verify that this variability
does not actually originate in one of the reference stars,
we repeated the analysis alternately using as
reference star \object{HD 157931} alone and the average
of the 8 other references. Both light curves are very
similar to Fig.~\ref{fig:phot1}.

The photometric observations are consistent with the signal of
a single spot, within the limitations of their incomplete phase
coverage: the variations are approximately sinusoidal,
and their $\sim$0.2-0.3~radian phase shift from the corresponding
radial velocity signal closely matches the difference expected
for a spot.
The spot would cover 2.6\% of the stellar surface if completely dark,
corresponding to a $\sim 0.16~R_\star$ radius for a circular spot.

\begin{figure*}
\centering
\includegraphics[width=0.9\linewidth]{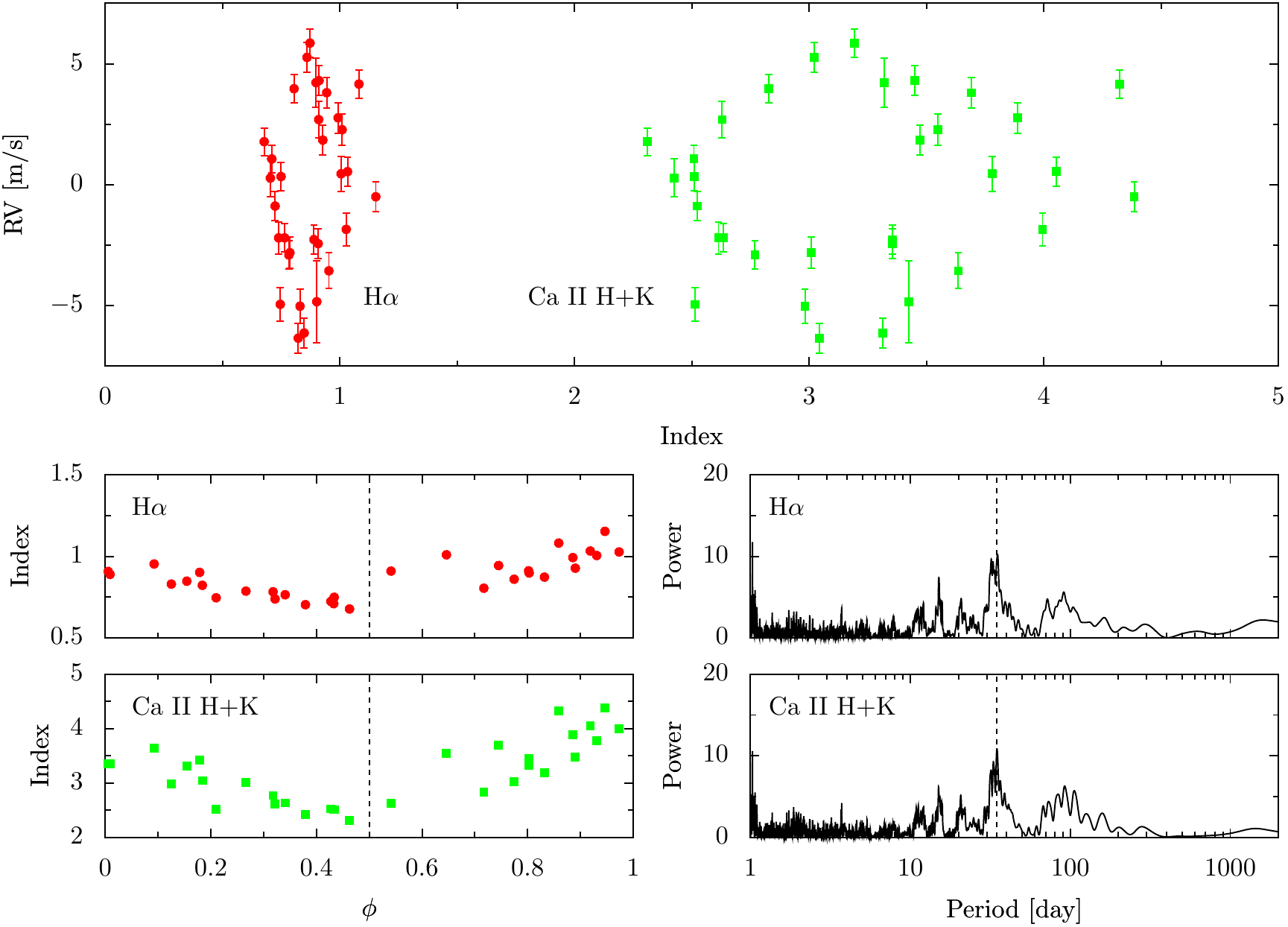}
       \caption{{\it Upper panel:} Differential radial velocity 
     of GJ 674, corrected for the signature of the 4.69~days planet
     in our 2-planet Keplerian fit, as a function of the H$\alpha$ 
     (red filled circles) and \ion{Ca}{ii} H\&K (green filled squares)
     spectral indices defined in the text. 
       {\it Bottom right panels:} the \ion{Ca}{ii} H+K and H$\alpha$ 
     indexes phased to the longer period of the 2-planet Keplerian 
     model. 
{\it Bottom left panels:} Power Density spectra of the spectroscopic
     indexes. A clear power excess peaks at 34.8 days (vertical 
     dashed lines).
       }
       \label{fig:index}
\end{figure*}

\subsection{\label{subsect:spec}Variability of the spectroscopic indices}
The emission reversal in the core of the \ion{Ca}{ii} H\&K resonant
lines results from non-radiative heating of the chromosphere, which
is closely coupled to spots and plages through magnetic connections
between the photosphere and chromosphere. The H$\alpha$ line 
is similarly sensitive to chromospheric activity. We measured these
chromospheric spectral features results in the clean HARPS
spectra used to measure the radial velocities, and examine
their variability.

Like the well known Mt. Wilson S index \citep{Baliunas1995},
our \ion{Ca}{ii} H$+$K index is defined as:
\begin{equation}
\mathrm{Index}=\frac{H+K}{B+V}.
\end{equation}
with $H$ and $K$ sampling the two lines of the \ion{Ca}{ii}
doublet, and  $B$ and $V$ the continuum on both sides of the doublet.
Our $H$ and $K$ intervals are 31~$\mathrm{km\,s^{-1}}$ wide
and centered on 3933.664 and 3968.47~\AA, while $B$ and $V$ are
respectively integrated over [3952.6, 3956 \AA] and
[3974.8, 3976~\AA].

This H$+$K index varies with a clear period of
$\sim$34.8 days (Fig.~\ref{fig:index}). Within
the combined errors this is consistent
with both the photometric period and the
longer radial velocity period. The phasing of
the chromospheric index and the photometry is such
that lower photometric flux matches
higher \ion{Ca}{ii} emission, as expected if active
chromospheric regions hover over photospheric spots. 

A plot of the (apparent) radial-velocity as a function 
of the  H$+$K spectral index similarly shows the 
characteristic loop pattern expected for a spot.
The radial velocity effect of a spot cancels out when 
it crosses the sub-observer meridian, which 
occurs twice during a rotation period: once on 
the hemisphere facing the observer, and once on the opposite
hemisphere. During the front-facing crossing the spot has 
maximal projected area, hence maximal chromospheric 
emission, while it has a minimal projected area
(and is possibly hidden, depending on its latitude 
and the stellar inclination) during the back-facing
crossing. As a result, both extrema of the chromospheric 
index correspond to radial-velocity zero-crossings.
At intermediate phases the spot produces intermediate
chromospheric emission levels, and it induces positive 
(respectively negative) radial-velocity shifts when the 
masked area is on the rotationally blue- (respectively 
red-shifted) half of the star. The net result in a 
plot of chromospheric emission as a function of radial 
velocity is a closed loop.

Chromospheric filling-in of photospheric H$\alpha$ 
absorption has similarly been found a powerful activity 
diagnostic for M dwarfs. \citet{Kurster2003} found that
in Barnard's star it correlates linearly with the 
radial-velocity variations, and interpreted that 
finding as evidence that active plage regions
inhibit the convective velocity field. 
The variation pattern in \object{GJ~674}
definitely differs from a linear correlation between 
H$\alpha$ and the radial-velocity residuals,
and needs a different explanation.

Similarly to \citet{Kurster2003} we define our
H$\alpha$ index as:
\begin{equation}
\mathrm{Index}=\frac{F_{H\alpha}}{F_1+F_2}.
\end{equation}
with $F_{H\alpha}$ sampling the H$\alpha$ line, and  $F_1$ 
and $F_2$ the continuum on both sides of the line. Our 
$F_{H\alpha}$ interval is 31~$\mathrm{km\,s^{-1}}$ wide
and centered on 6562.808~\AA, while $F_1$ and $F_2$ are
respectively integrated over [6545.495, 6556.245~\AA] 
and [6575.934, 6584.684~\AA]. The H$\alpha$ index behaves 
similarly to the \ion{Ca}{ii} H+K index.

The chromospheric indices vary by factors of $\sim$2
and $\sim$1.3 (for our specific choices of continuum
windows), and are thus much more contrasted than 
the photometry. They do not however vary as smoothly 
with phase as the photometry, perhaps due to 
(micro-)flares. This somewhat reduces their value
as diagnostics of spot-induced radial velocity 
variations, but these measurements on the other 
hand require no new observation. They undoubtedly 
reinforce the spot interpretation here, and they will
be extremely useful in cases where photometry cannot
be immediately obtained.

\subsection{\label{sect:blah}Planets vs. activity}

In \S\ref{sect:orbit} we showed that our 32
radial-velocity measurements of \object{GJ~674} are
well described by two Keplerian signals, as
illustrated by the low reduced chi-square of that
model. The above analysis (\S\ref{sect:activity})
however demonstrates that the rotation period of
\object{GJ 674} coincides with the longer of the
two Keplerian periods. Both the stellar flux and
the \ion{Ca}{ii} H$+$K emission vary with that
period, implying that the surface of \object{GJ~674}
has a magnetic spot. This spot must induce
radial-velocity changes, with the observed phase
relative to the photometric signal. As a consequence,
some, and probably all, of the 35-day radial-velocity signal
must originate in the spot. Planet-induced activity
through magnetic coupling \citep[e.g.~][]{Shkolnik2005}
would in principle be an alternative explanation of
the correlation, but here it is not a very attractive
one: the inner planet is at least as massive as the
hypothetical 35-day planet, and would, at least
naively, be expected to have stronger interactions
with the magnetosphere of \object{GJ 674}. The
4.69-day period however is only seen in the radial velocity
signal, and it has no photometric or chromospheric
counterpart.

\section{Discussion}
\subsection{Characteristics of \object{GJ~674b}}
Perhaps the most important result of the above analysis is
that the $\sim$4.69-day planet of \object{GJ~674} is robust:
variability identifies the stellar rotation period as
$\sim$35~days, and the 4.69-day period therefore cannot
reflect rotation modulation. The short period signal,
in spite of its larger amplitude, also has no counterpart
in either photometry or chromospheric emission, further
excluding a signal caused by magnetic activity.

The 1-planet fit, which effectively treats the activity
signal as white noise, results in a minimum mass for
\object{GJ~674b} of $m_2 \sin i~=~12.7~\Mearth$.
The 2-planet fit by contrast filters out this signal.
That filtering obviously uses a physical model which
is not completely appropriate, but that remains
preferable to handling a (partly) coherent signal as
white noise. We therefore adopt the corresponding
estimate of the minimum mass, $m_2~\sin~i~=~11.09~\Mearth$.

At 0.039 AU from its parent star, the temperature
of \object{GJ~674~b} is $\sim$450~K. Planets above a few
Earth masses planets can, but need not, accrete a large
gas fraction, leaving its composition -- mostly gaseous or
mostly rocky -- unclear. The orbital eccentricity might
shed light on the structure of \object{GJ~674b}, if confirmed
by additional measurements: rocky and gaseous planets have
rather different dissipation properties, and significant
eccentricity at the short period of \object{GJ~674~b}
needs a high Q factor, unless it is pumped by an additional
planet at a longer period \citep[e.g. ][]{Adams2006}. For 
now, the stellar activity leaves the statistical significance 
of the eccentricity slightly uncertain, and we therefore 
prefer to stay clear from overinterpreting it.

\subsection{Properties of M-dwarf planets}
One important motivation in searching for planets around
M dwarfs is to investigate whether the planet-metallicity
correlation found for Jupiter-mass planets around solar-type
stars extends to very-low-mass stars. Our photometric
calibration of M dwarfs metallicity \citep{Bonfils2005b}
gives respective metallicities of [Fe/H]=$-0.03$, $-0.25$,
$+0.14$, $+0.03$, and $-0.28$ for \object{GJ 436},
\object{GJ 581}, \object{GJ 849}, \object{GJ 876}
and \object{GJ~674}. M dwarfs with known planets
therefore have an average metallicity of $-0.078$
and a median of $-0.03$. By comparison, the 44
M dwarfs of the \citet{Bonfils2005b} volume limited
sample which are not currently known to host a planet
have average and median metallicities of $-0.181$
and $-0.160$. M dwarfs with planets therefore appear
slightly more metal-rich than M dwarfs without planets.
A Kolmogorov-Smirnov test \citep{Press1992} of the two
samples gives an 11\% probability that they are drawn
from the same distribution.
The significance of the discrepancy is therefore still 
modest, limited by small-number statistics.

One can additionally note that the two stars which
host giant planets, \object{GJ 876} and \object{GJ 849},
occupy the metal-rich tail of the M dwarf metallicity
distribution, with \object{GJ 849} almost as metal-rich
as the most metal-rich star of the comparison sample.
The next most metal-rich of the M dwarfs with planets,
\object{GJ 436}, has an additional long-period companion
(P$>$6~yr) which might well be a giant planet
\citep{Maness2006} and would then strengthen that trend.
If confirmed by additional data, this would validate
the theoretical predictions \citep{Ida2004, Benz2006}
that only Jovian-mass planets are more likely to form
around metal-rich stars. Current observations are
consistent with this prediction, but not yet very
conclusively so \citep{Udry2006}.

\begin{figure}
\centering
\includegraphics[width=0.9\linewidth]{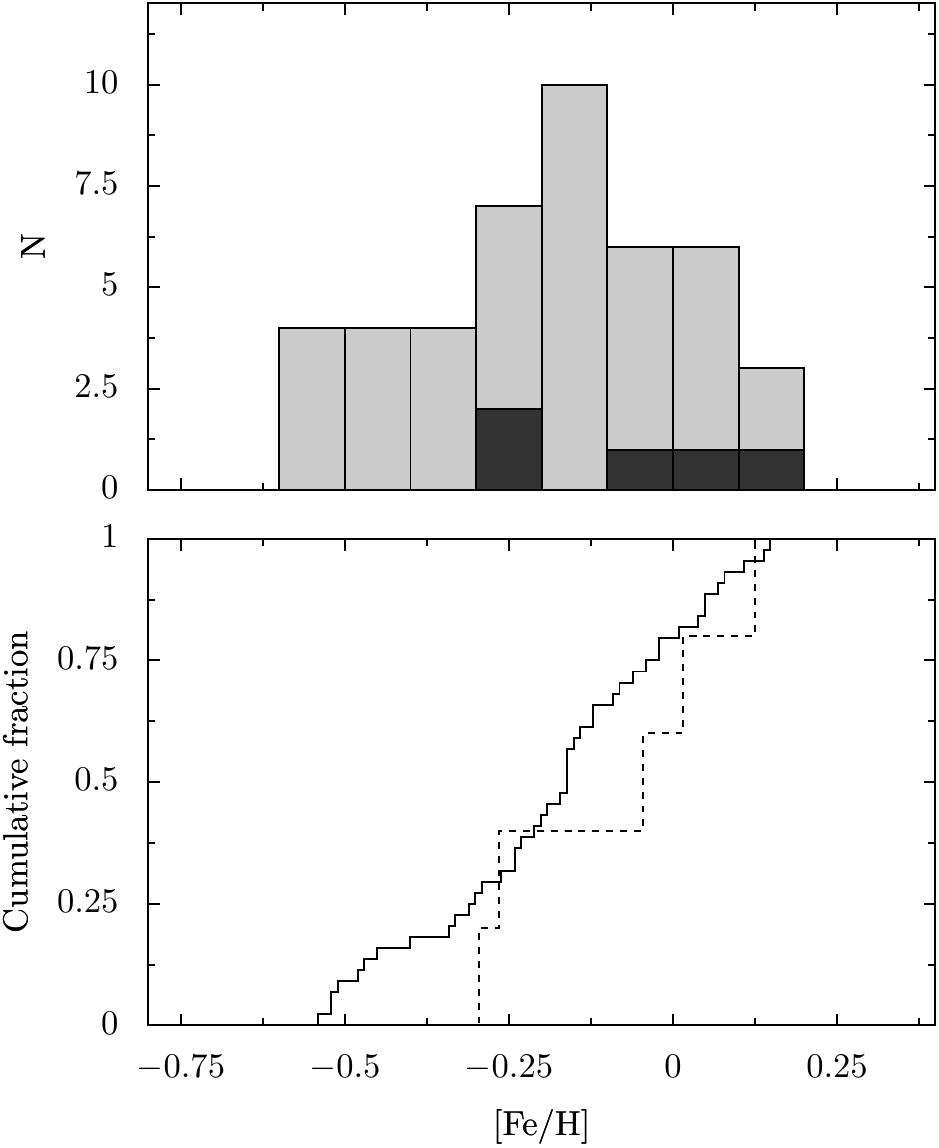}
       \caption{{\it Upper panel:} Metallicity distributions of 
    44~M dwarfs without known planets (gray shading) and of the 5~M 
    dwarfs known to host planets (black shading).
      {\it Bottom panel:} Corresponding cumulative distributions
      (solid and dashed lines, respectively).
       }
       \label{fig:feh}
\end{figure}

Much recent theoretical work has gone into examining how
planet formation depends on stellar mass. Within the
``core accretion'' paradigm, \citet{Laughlin2004}
and \citet{Ida2005} predict that giant planet formation
is inhibited around very-low-mass stars, while Neptune-mass
planets should inversely be common. Within the same paradigm,
but assuming that M dwarfs have denser protoplanetary
disks, \citet{Kornet2006} predict instead that
Jupiter-mass planets become more frequent in inverse
proportion to the stellar mass. Finally,
\cite{Boss2006a} examines how planet formation depends
on stellar mass for planets formed by disk instability,
and concludes that frequency of Jupiter-mass planet is
independent of stellar mass, as long as disks are massive
enough to become unstable.

To date, none of the $\sim$300 M dwarfs scrutinized
for planets by the various radial-velocity searches
\citep{Bonfils2006, Endl2006, Butler2006} has
been found to host a hot Jupiter.
Conversely, \object{GJ~674b} is already the
4$^{th}$ hot Neptune. Though that cannot be
established quantitatively yet, these surveys
are likely to be almost complete for hot Jupiters,
which are easily detected. Hot Neptune detection,
on the other hand, is definitely highly incomplete.
Setting aside this incompleteness for now,
simple binomial statistics shows that the probability
of finding no and 4 detections in 300 draws of the same
function is only 3\%. There is a thus 97\% probability
that hot Neptunes are more frequent than hot Jupiter
around M dwarfs. Accounting for this detection bias
in more realistic simulations (Bonfils et al. in prep.)
obviously increases the significance of the difference.
Planet statistics around M dwarfs therefore favor
the theoretical models which, at short periods,
predict more Neptune-mass planets than Jupiter-mass
planets.

\begin{table}
\caption{Keplerian parameterization for GJ~674b and GJ~674's spot.}
\label{orbparam}
\centering
\begin{tabular}{l l r@{~$\pm$~}l r@{~$\pm$~}l}
\hline
\multicolumn{2}{l}{\bf Parameter} & \multicolumn{2}{c}{\bf GJ~674b} & \multicolumn{2}{c}{\bf Spot}\\
\hline
$P$                 & [days]                 & 4.6938		& 0.007	& 34.8467		& 0.0324\\
$T$                 & [JD]                     & 2453780.085	& 0.078	& 2453767.13	& 0.92 \\
$e$                 &                             & 0.20		& 0.02       	& 0.20 		& 0.05\\
$\omega$      & [deg]                  & 143 		& 6 		& 113		& 9\\
$K$                 & [m s$^{-1}$]      & 8.70		& 0.19 	& 5.06		& 0.19\\
\hline
$a_1 \sin{i}$  & [AU] & \multicolumn{2}{c}{3.68 10$^{-6}$}		& \multicolumn{2}{c}{1.59 10$^{-5}$} \\
$f(m)$             & [M$_{\odot}$] & \multicolumn{2}{c}{3.0 10$^{-13}$}& \multicolumn{2}{c}{4.4 10$^{-13}$} \\
$m_2 \sin{i}$ & [$M_{\mathrm{Earth}}$] & \multicolumn{2}{c}{11.09}& \multicolumn{2}{c}{12.58}\\
$a$ & [AU]     & \multicolumn{2}{c}{0.039}						& \multicolumn{2}{c}{0.147}\\
\hline
\end{tabular}
\end{table}

\begin{acknowledgements}
We are grateful to the anonymous referee for constructive 
comments. XB and NCS acknowledge support from the Funda\c{c}\~ao para
a Ci\^encia e a Tecnologia (Portugal) in the form of fellowships
(references SFRH/BPD/21710/2005 and SFRH/BPD/8116/2002) and a 
grant (reference POCI/CTE-AST/56453/2004). The photometric 
monitoring has been performed on the EULER 1.2 meter telescope 
at La Silla Observatory. We are grateful to the SNF 
(Switzerland) for its continuous support. This research has
made use of the SIMBAD database, operated at CDS, Strasbourg, France.
\end{acknowledgements}

\bibliographystyle{aa}
\bibliography{maBiblio}

\appendix
\section{\label{subsect:bis}Bisector analysis}
\begin{figure}
\centering
\includegraphics[width=0.9\linewidth]{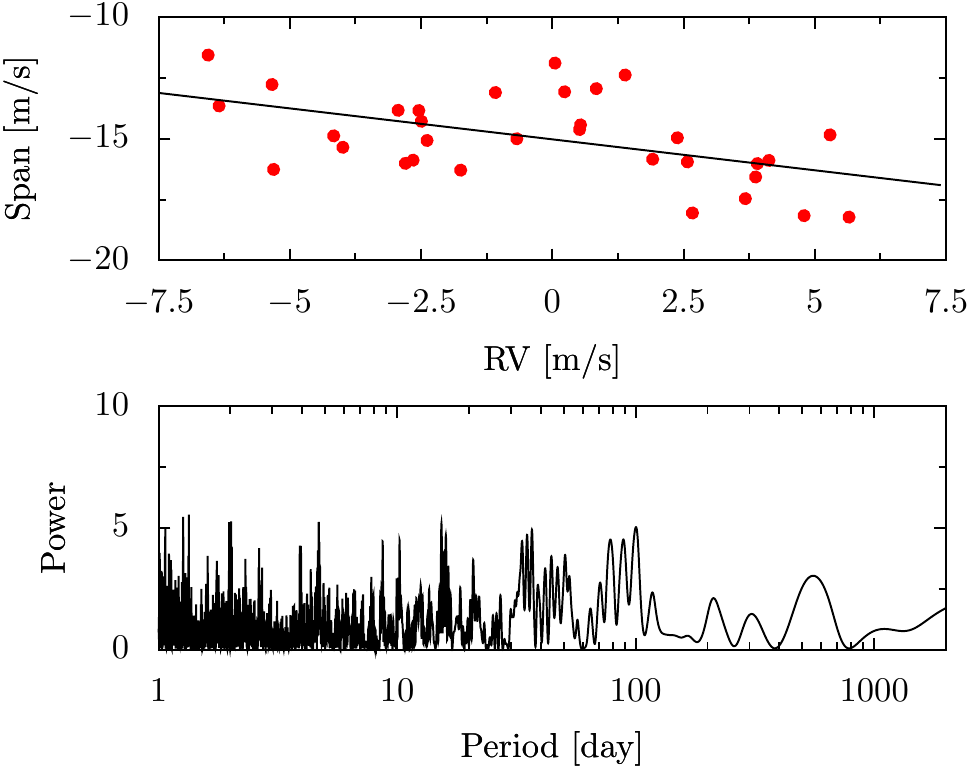}
       \caption{Bisector analysis for GJ 674 measurements
       }
       \label{fig:bis}
\end{figure}

As demonstrated by \citet{Saar1997} the bisector analysis
loses much of its diagnostic power when applied to slow 
rotators. In simulations of the impact of star spots 
on radial-velocity and bisector measurements, they found 
that, for a given spot configuration, the radial velocity 
varies linearly with $v \sin i$ while the bisector span 
varies as $(v \sin i)^{3.3}$. 
The bisector signal therefore decreases faster with decreasing 
rotational velocities than the radial-velocity signal, and disappears
faster in measurement noise. For \object{GJ 674} we measure a 
very low rotation velocity ($v~\sin~i < 1 \mathrm{km\,s^{-1}}$). 
It is therefore unsurprising that the correlation between the 
bisector span and radial velocity is weak (Fig.~\ref{fig:bis})
and not statistically significant.
\end{document}